\documentclass{aastex631}

\usepackage{graphicx}

\begin{document}

\title{Comparing Optical Variability of Type 1 and Type 2 AGN from the BAT 9-Month Sample using ASAS-SN and TESS Surveys}

\author{Natalie Kovacevic}
\affiliation{Homer L.\ Dodge Department of Physics and Astronomy,
University of Oklahoma, Norman, OK 73019, USA}
\email{natalie.kovacevic@ou.edu}

\author[0000-0001-9203-2808]{Xinyu Dai}
\affil{Homer L.\ Dodge Department of Physics and Astronomy,
University of Oklahoma, Norman, OK 73019, USA}
\email{xdai@ou.edu}

\author[0000-0002-7720-3418]{Heechan Yuk}
\affil{Homer L.\ Dodge Department of Physics and Astronomy,
University of Oklahoma, Norman, OK 73019, USA}

\author{Emilia E.\ Jaevelae}
\affiliation{Department of Physics and Astronomy, Texas Tech University, Box 41051, Lubbock, TX, 79409-1051, USA}
\affiliation{Homer L.\ Dodge Department of Physics and Astronomy,
University of Oklahoma, Norman, OK 73019, USA}

\author[0000-0001-8920-0073]{Tingfeng Yi}
\affiliation{Key Laboratory of Colleges and Universities in Yunnan Province for High-energy Astrophysics, Department of Physics, Yunnan Normal University, Kunming 650500, China}

\author[0000-0001-5661-7155]{Patrick J. Vallely}
\altaffiliation{NSF Graduate Research Fellow}
\affiliation{Department of Astronomy, The Ohio State University, 140 West 18th Avenue, Columbus, OH 43210, USA}

\author[0000-0003-4631-1149]{Benjamin J. Shappee}
\affiliation{Institute for Astronomy, University of Hawai\`{}i at Manoa, 2680 Woodlawn Dr., Honolulu, HI 96822, USA}

\author[0000-0001-8973-5051]{Francesco Shankar}
\affiliation{School of Physics and Astronomy, University of Southampton, Highfield, Southampton 1BJ, UK}
\author{K.~Z.~Stanek}
\affiliation{Department of Astronomy, The Ohio State University, 140 West 18th Avenue, Columbus, OH 43210, USA}
\affiliation{Center for Cosmology and AstroParticle Physics, The Ohio State University, 191 W.\ Woodruff Ave., Columbus, OH 43210, USA}

\begin{abstract}
We present an optical variability analysis and comparison of the samples of Seyfert 1 and 2 galaxies, selected from the \textit{Swift} 9-month BAT catalog, using the light curves from Transiting Exoplanet Survey Satellite (TESS) and All-Sky Automated Survey for SuperNovae (ASAS-SN). 
We measured the normalized excess variance of TESS and ASAS-SN light curves for each target and performed a Kolmogorov-Smirnov test between the two samples, where our results showed significant differences. This is consistent with predictions from the unification model, where Seyfert 2s are obscured by the larger scale dust torus and their variability is suppressed. 
This variability difference is independent of the luminosity, Eddington ratio, or black hole mass, further supporting geometrical unification models.
We searched the dependence of the normalized excess variance of Seyfert 1s on absolute magnitudes, Eddington ratio, and black hole mass, where our results are consistent with relations found in the literature. Finally, a small sub-sample of changing-look AGNs that transitioned during the time frame of the ASAS-SN light curves, with their variability amplitudes changing according to the classification, larger variability as type 1s and smaller as 2s.
The change of variability amplitudes can be used to better pinpoint when the type transition occurred.  
The consistency trend of the variability amplitude differences between Seyfert 1s and 2s and between changing-look AGNs in 1 or 2 stages suggests that variability can be a key factor in shedding light on the changing-look AGN or the dichotomy between Seyfert 1 or 2 populations.

\end{abstract}


\section{Introduction} \label{sec:intro}

Active galactic nuclei (AGNs) are luminous sources in our universe powered by the accretion of material onto a central supermassive black hole. The unification model (\citealt{Antonucci_1993ARA&A..31..473A,Urry_1995PASP..107..803U}) was proposed to classify these AGNs where the model consists of a broad line region (BLR) around the disk, a dusty torus, a much larger narrow line region (NLR), and a relativistic jet. The viewing angle of the AGN will determine the type classification. Type 1s have a clear view of the central engine, and thus their optical spectra show both broad and narrow emission lines. Type 2s have an obscured view of the central black hole due to the torus, exhibiting only narrow emission lines in their spectra. There are also sub-classes of AGNs (i.e., quasar, Seyfert, blazar) with Seyferts being less luminous than quasars and blazars showing evidence of a relativistic jet towards the observer. Within the Seyfert 1 (Sy1) and Seyfert 2 (Sy2) classifications, there are further sub-classes: the intermediate types. Type 1.2, 1.5, and 1.8 refers to the relative strength of the broad H$\beta$ component, the strength of the broad component decreases as the type classification increases. A type 1.9 has no broad H$\beta$ component but does have a broad H$\alpha$ component.
Recently, some Seyferts are found to transition between AGN sub-types as changing-look AGNs.

A key feature of AGNs is their variability across all wavelength bands. A number of studies have been carried out on the variability of type 1 AGNs, but few studies have focused on the variability of type 2 AGNs. In the X-ray band, type 1 and type 2 AGNs have been found to vary in both the short timescale (hours) and long timescale (years) from examining light curves detected with XMM-Newton \citep{mateos_2007A&A...473..105M}. From the optical broad-band monitoring campaign of 35 Seyfert galaxies including both type 1s and type 2s, \citet{Winkler_1992MNRAS.257..659W} noticed that most galaxies in their sample were variable. However, the optical emission lines of type 2s generally have a near-negligible variability as seen in \citet{Yip_2009AJ....137.5120Y}, further explaining that the host galaxy is dominating the optical continuum flux. \citet{Sánchez_2017} found that observing at longer wavelengths from the optical to the NIR decreased the fraction of variable type 1 and type 2 AGNs, suggesting that in the NIR reprocessed emission from the dusty torus can be seen and damped variability can be detected. Other studies on type 2 variability include \citet{choi_2014ApJ...782...37C}, who found that two out of six of their type 2s showed non-negligible variability on long timescales from examining Sloan Digital Sky Survey (SDSS) data, and \citet{Cartier_2015ApJ...810..164C} who examined a small sample of data from the QUEST-La Silla AGN variability survey classified 21 percent of their narrow-line sample as variable. Many mechanisms have been proposed to explain this variability including, thermal fluctuations \citep{Kelly_2009}, accretion disk instabilities (\citealt{Kawaguchi_1998, Trevese_2002}), and supernova explosions in host galaxies or accretion disk \citep{Aretxaga_10.1093/mnras/286.2.271}.

In recent years, a new phenomenon has been reported: the Changing-look (CL) AGN. A CL AGN can be observed in the X-ray when the X-ray absorption changes from Compton-thick to Compton-thin \citep{Matt_2003MNRAS.342..422M} or vice-versa and in optical spectroscopy where the AGN changes from a type 1 classification to a type 2 or vice-versa. A CL AGN can also transit from one sub-type to another in the case of blazars \citep{Mishra_2021}. In the optical this has been reported in Seyferts (\citealt{Tohline_1976ApJ...210L.117T,Denney_2014ApJ...796..134D}) and quasars \citep{LaMassa_2015ApJ...800..144L}). 
This is a relatively new field of AGN research and the driving mechanism for this phenomenon is under debate mainly between drastic variations to the accretion rate \citep{Stern_2018} and variable obscuration \citep{Elitzur_2012ApJ...747L..33E}, among other proposed ideas, for instance tidal disruption events \citep{Merloni_2015MNRAS.452...69M}. 

In this paper, we examine and compare the variability between Sy1s and Sy2s using two different datasets in an effort to establish a clear distinction in their variability levels and apply this result to the optical light curves of CL AGNs.
In Section \ref{Data}, we discuss the parent sample for our study and introduce the Transiting Exoplanet Survey Satellite (TESS) and All-Sky Automated Survey for SuperNovae (ASAS-SN) data and light curve extraction process. In Section \ref{Analysis} we present our method for comparing the variability using the normalized excess variance. In Section \ref{Results} we present how the variability of Sy1s compares to Sy2s and show how the variability is dependent on the absolute magnitude, black hole mass, and Eddington ratio. We also highlight an interesting Sy2 that exhibited above normal variability and finally using the results of our comparison to constrain the transition time for the few CL AGNs in our sample. In Section \ref{conclusion} we summarize our results. Throughout the paper we assume cosmological parameters of H$_0$ = 72 kms$^{-1}$Mpc$^{-1}$, $\Omega_m$ = 0.3, and $\Omega_\Lambda$ = 0.7.

\section{Sample Selection, Data, and Calibrations} \label{Data}

\subsection{\textit{Swift}-BAT 9-Month Survey} \label{sample}
We focus our study on the sample of AGNs from the first 9 months of the hard X-ray all-sky survey \citep{Tueller_2008ApJ...681..113T} using the Burst Alert Telescope \citep[BAT;][]{Barthelmy_2005SSRv..120..143B} on the Neil Gehrels \textit{Swift} Observatory
\citep{Gehrels_2004ApJ...611.1005G}.
This catalog contains mostly bright AGNs, which match well with the optical survey characteristics of ASAS-SN and TESS, such that well sampled and high S/N light curves are available for almost all of the sources.
Later BAT AGN catalogs \citep[e.g.,][]{Oh_2018ApJS..235....4O} contain more fainter sources, which we will explore in future studies.
The first 9-month catalog contains AGNs with the classification of blazar, BL Lac, galaxy, Low-ionization nuclear emission-line region (LINER), Sy1, Sy2, and intermediate Seyfert types. For our sample we have excluded seven blazars, 10 BL Lacs, six galaxies, and one LINER from the BAT catalog, focusing our analysis sample on the Sy1, 2, and intermediate types. The final selected sample contains 108 AGNs divided into 55 Sy1s and 53 Sy2s, making the sample ideal to compare their variability characteristics. We further grouped Sy1, 1.2, and 1.5 as unobscured AGNs and Sy1.8, 1.9, and 2 sources as obscured AGNs in accordance to the strength/lack of a broad H$\beta$ component \citep{Osterbrock_1981ApJ...249..462O}.

It is important to identify jetted AGNs when analyzing the optical variability of the sample since the relativistic jet provides an additional source of variability.  We have already excluded blazars in the sample selection, and we further identify jetted AGNs by their radio loudness $F_{5GHz}$/$F_B$ $>$ 10 \citep{Chiaberge_2011MNRAS.416..917C}, where $F_B$ is their B-band flux and their radio fluxes were taken from the NASA/IPAC Extragalactic Database (NED)\footnote{The NED website is at https://doi.org/10.26132/ned1.}. We identified seven out of 55 Sy1s and 11 out of 53 Sy2s as radio loud. 
A follow-up literature review was conducted to confirm that the sources meeting the criteria to be radio loud do in fact have a radio jet and radio quiets have no indication of a jet. 
We have also identified two narrow-line Sy1s in the process, Mrk 110 and IGRJ21277+5656 (\citealt{halpern_2006ATel..847....1H, veron_2006A&A...455..773V}), which satisfy the criteria of having a FWHM($H_\beta$) $<$ 2000 kms$^{-1}$ and a flux ratio of [O III]$\lambda$5007 / $H_\beta$ $<$ 3 (\citealt{Osterbrock_1985ApJ...297..166O,Goodrich_1989ApJ...342..224G}). These have been excluded due to Mrk 110 displaying a higher than average variability.

\subsection{ASAS-SN and TESS Light Curves} \label{sec:light curves}
Data for our sample was obtained from the Transiting Exoplanet Survey Satellite (TESS) and All-Sky Automated Survey for SuperNovae (ASAS-SN). TESS is an all-sky survey and its main goal of the 2 year primary and subsequent extended missions was to identify transiting exoplanets around M dwarf stars \citep{2020AJ....160..116G}. However, the precision and high cadence capabilities of TESS, make it a useful tool for studying the variability of AGNs on a shorter timescale (\citealt{2023MNRAS.525.5795T,Yuk_2023arXiv230617334Y}). We used full frame images (FFIs) for our sample of AGNs in both the TESS primary mission (sectors 1--26) and its first extended mission (sectors 27--55). Each sector comprises of two cycles each with approximately 13 days for a total light curve length of approximately 27 days. 
 The FFIs were taken every 30 min in its primary mission and taken every 10 min in the first extended mission.

\begin{figure}
\includegraphics[width=1.0\textwidth, trim={0cm 1cm 0 2cm}]{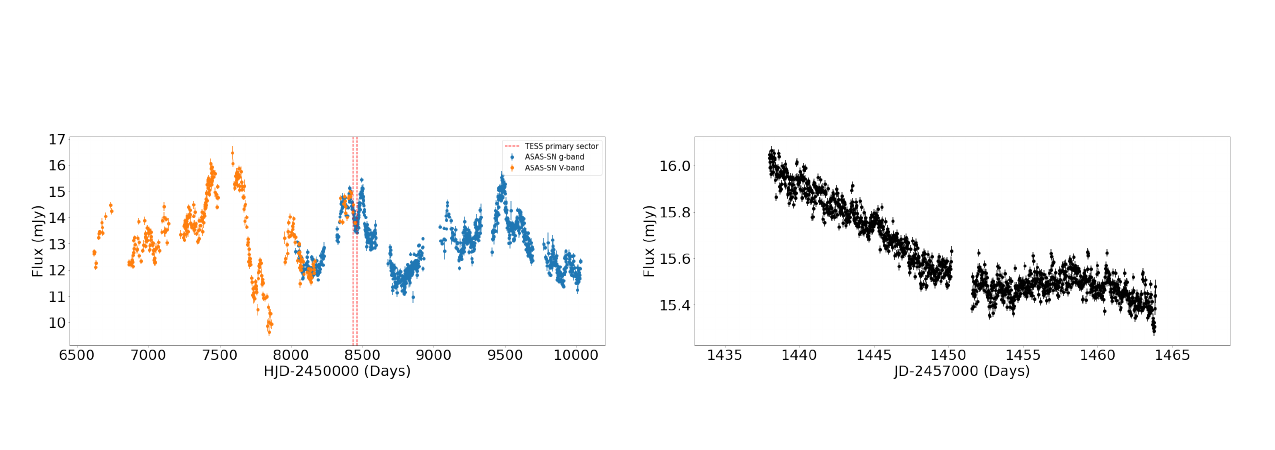}
\caption{Left panel: ASAS-SN light curve of the Sy1 Ark 120 showing the flux in both V-band and g-band (the V and g light curves are normalized) as an example. The visual variability of the type 1 can be seen in the ASAS-SN light curve by stochastic variations. The red dotted lines indicate the time frame the TESS light curve takes place. Right panel: TESS light curve showing this stochastic variability on a shorter timescale.
\label{type1}}
\end{figure}

\begin{figure}
\includegraphics[width=1.0\textwidth, trim={0 0 0 2cm}]{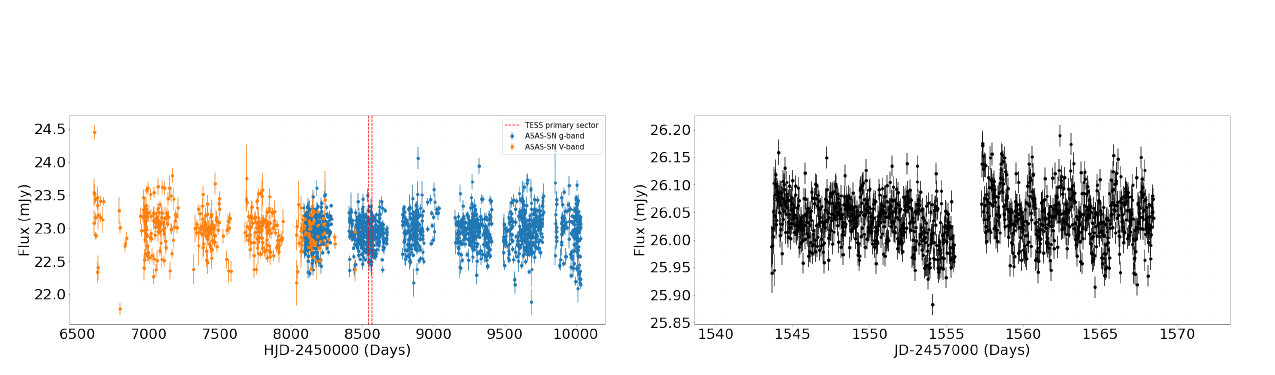}
\caption{Left panel: ASAS-SN light curve showing the flux in both V-band and g-band (the V and g light curves are normalized) of the Sy2 NGC 3081 as an example. The red dotted lines indicate the time frame the TESS light curve takes place. Visually the type 2 has little variability beyond the noise in the ASAS-SN light curve. Right panel: TESS light curve of the Sy2 NGC 3081.
The type 2 light curves exhibit smaller amplitudes of variability.
\label{type2}}
\end{figure}

The process used for extracting the TESS light curves is described in \citet{2021MNRAS.500.5639V}, where the authors developed an automated image subtraction pipeline. This pipeline utilizes four subtraction methods; the basic ISIS subtraction, median filtering, Gaussian smearing, and both the Gaussian smearing and filtering. The median filtering method incorporates a background filtering that uses one-dimensional median filters which are applied along each axis of the CCD, following the initial ISIS subtraction. This process helps alleviate the issue of CCD strap artifacts found in many TESS observations. Based on visual inspection for systematical effects, the light curves produced from median filtering are the most appropriate to use in our case for further analysis. For light curves that exhibited potential systematical effects, the light curves of the nearby pixels were examined to determine the origin of the anomalies. One such origin is from nearby variable stars, since \citet{Treiber_2023MNRAS.525.5795T} flagged more than 30\% of their TESS light curves as being contaminated by nearby variable stars. However, due to the brightness difference between the sample of the first 9 months of the BAT catalog ($V<16$) used in this paper compared to the \citet{Treiber_2023MNRAS.525.5795T} sample ($I<20$), the variable star contamination had a much smaller effect, affecting 3\% of our sources. We determined three sources that were contaminated by variable stars, and these sources were removed from our final sample for analysis.
Examples of these light curves for both a typical Sy1 and Sy2 can be seen in the right panels of Figure~\ref{type1} and Figure~\ref{type2}. The FFIs used in this work were obtained from the Mikulski Archive for Space Telescopes (MAST)\footnote{The MAST website is at https://doi.org/10.17909/0cp4-2j79}. 

ASAS-SN, in contrast, is a suite of telescopes that has been observing the entire visible sky every night since 2017. The primary goal of ASAS-SN is to carry out a survey for bright transients (\citealt{Shappee_2014ApJ...788...48S, 2017PASP..129j4502K}). Although ASAS-SN has been operating since 2012, it originally had 2 mounts that used the V-band filter and in 2017 was upgraded to 5 mounts. With these 5 mounts, 3 of them used the g-band filter while the original 2 still used the V-band filter. Then in 2018, the original 2 mounts were changed to the g-band filter \citep{2022ApJ...930..110Y}, resulting in an approximate 400 day overlap of V-band and g-band in the ASAS-SN light curves where we can normalize the V and g-band light curves. The ASAS-SN light curves are generated from a variant of the image subtraction method and are calibrated against the ensemble of stars in the observing fields (\citealt{jayasinghe_2018MNRAS.477.3145J, jayasinghe_10.1093/mnras/stz444}).
Examples of these light curves for Sy1 and Sy2 can be seen in the left panel of Figure~\ref{type1} and Figure~\ref{type2}  with the V band spanning approximately 1500 days and the g band spanning approximately 2050 days for a total light curve length of approximately 3250 days. The vertical red dotted lines indicate the time frame when the TESS light curve took place for that same source. With this, we are able to use ASAS-SN light curves to calibrate the flux of our TESS light curves. Having the long timescale ASAS-SN data is useful for comparing the results of the short timescale TESS data.

\section{Analysis: Normalized Excess Variance} \label{Analysis}

\begin{figure}
\includegraphics[width=1.0\textwidth, trim={0.5cm 0.5cm 0.5cm 3cm}]{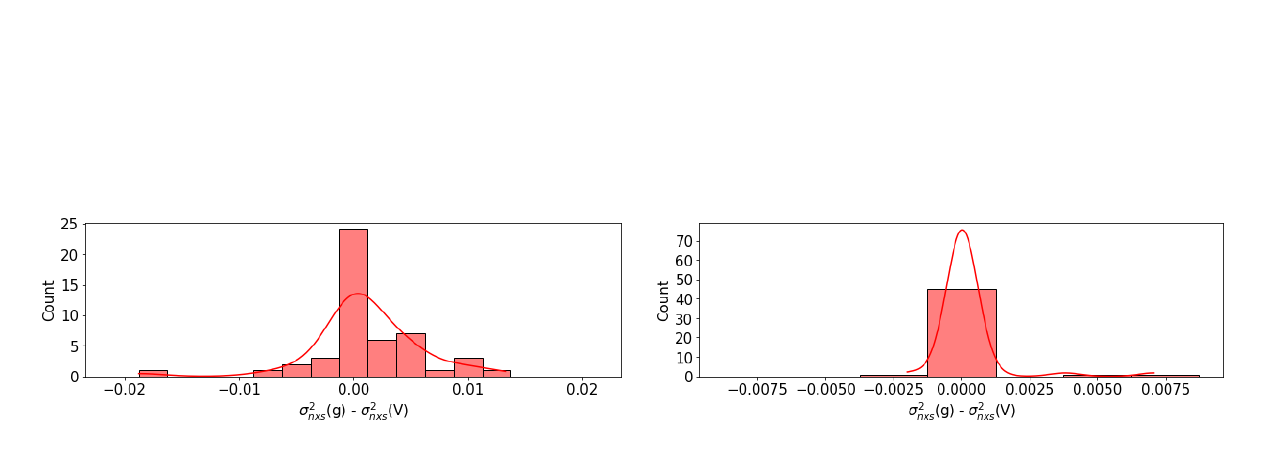}
\caption{Difference in $\sigma_{\textrm{nxs}}^2$ when calculated in the g-band and V-band ASAS-SN light curves, separately. When modeled using a Gaussian distribution, it peaks close to zero signifying that there is no significant offset between the variability measured in the two bands. Left: $\sigma_{\textrm{nxs}}^2$ difference using the Sy1 population. Right: $\sigma_{\textrm{nxs}}^2$ difference using the Sy2 population.
\label{nxs}}
\end{figure}

We used the normalized excess variance, the noise removed variance normalized by the mean squared of the source, to measure the variability of the sample. This is a commonly used metric to measure the intrinsic variability amplitude of the source (\citealt{ 2003MNRAS.345.1271V, 2019Galax...7...62S}). We chose to use the flux units when calculating the normalized excess variance for this sample via Equation~\ref{equation1} and performed the calculations for the ASAS-SN and TESS datasets separately, where $S^2$ represents the variance, $\sigma_{\textrm{err}}$ the uncertainty of the light curve, and $\langle{x}\rangle$ the mean of the light curve.

\begin{equation}\label{equation1}
\sigma_{\textrm{nxs}}^2 = \frac{S^2 - \langle{\sigma_{\textrm{err}}^2 }\rangle}{\langle{x}\rangle^2}
\end{equation}
The error in the normalized excess variance, which only accounts for the errors on the flux measurements, is 
\begin{equation}\label{equation2}
\Delta{\sigma_{\textrm{nxs}}^2} = \sqrt{\Biggl( \sqrt{\frac{{2}}{N}}\frac{\langle{\sigma_{\textrm{err}}^2 }\rangle}{\langle{x}\rangle^2} \Biggr)^2 + \Biggl({\sqrt{\frac{\langle{\sigma_{\textrm{err}}^2 }\rangle}{N}}\frac{2F_{\textrm{var}}}{\langle{x}\rangle}}\Biggr)^2}
\end{equation}
where $N$ is the number of data points and $F_{var}$ is the fractional variability.
\begin{equation}\label{equation3}
F_{\textrm{var}} = \sqrt{\sigma_{\textrm{nxs}}^2}
\end{equation}
The normalized excess variance was calculated using the combined, normalized $V$ and $g$-band light curves of the ASAS-SN data set.  Although the $g$-band filter is slightly bluer than the $V$-band, which can be subject to color-dependent variability when combining their light curves, the two bands are very close compared to the broad UV-optical-IR emission of the accretion disk.  We calculated the normalized excess variance of the more variable Sy1s and the less variable Sy2s in the $V$ and $g$ bands separately and compared their differences. We find the differences can be modeled by Gaussian distributions peaking close to zero (Figure \ref{nxs}). This suggests that there is minimal detectable color-dependent variability and both $V$ and $g$-band light curves are realizations of the approximately same random process. We choose to present the analysis based on the combined $V$ and $g$ light curves to increase the signal-to-noise ratios of the analysis and a simplified presentation of the analysis results.

It was evident in our analysis that the uncertainty for the ASAS-SN data was underestimated. This was corrected by utilizing the power spectral density of the light curve, which provides a more accurate measurement of the white noise contribution. The process involved measuring the total power of the light curve and the integral of the white noise constant that was produced by the Lomb-Scargle periodogram, taking the ratio of them, and adjusting the noise term in the normalized excess variance (Equation~\ref{equation1}). The normalized excess variance obtained by this calculation has taken into account the white noise contribution in the light curves.
We defer the detailed analysis of the power spectra or structure functions of the sample in a subsequent paper.

We found further measurement biases in the normalized excess variance by examining the median apparent magnitude for the ASAS-SN and TESS light curves where we observed an increase in $\sigma_{\textrm{nxs}}^2$ measurements going towards fainter Sy2s.
This is unexpected since the apparent magnitude is an arbitrary property and therefore it is highly unlikely that this trend should exist, and we attribute this anomaly as an additional measurement bias of underestimating the flux uncertainties as the source apparent magnitudes are fainter. A correction is applied by compensating the flux uncertainties for all sources such that there is no dependence of $\sigma_{\textrm{nxs}}^2$ values with the apparent magnitudes of Sy2s.

The majority of ASAS-SN light curves are roughly evenly sampled with gaps between observational seasons. We use Equation 11 presented in \citet{Allevato_2013ApJ...771....9A} to correct for these seasonal gaps. However, with a PSD slope of $\beta$ = 1 \citep{Yuk_2023arXiv230617334Y} no bias correction was necessary, with a correction factor of 1.  
The TESS light curves have roughly a one day cycle gap (a small amount of TESS light curves are unevenly sampled), again using Equation 11 from \citet{Allevato_2013ApJ...771....9A} to correct for this cycle gap. With a PSD slope of $\beta$ = 2 \citep{Yuk_2023arXiv230617334Y}, a bias correction was applied.

\section{Results}
\label{Results}
\subsection{Dependence on Absolute Magnitude, Eddington Ratio, and Black Hole Mass} 

\begin{figure}[ht]
\includegraphics[width=\textwidth, trim={0 0 0 3cm}]{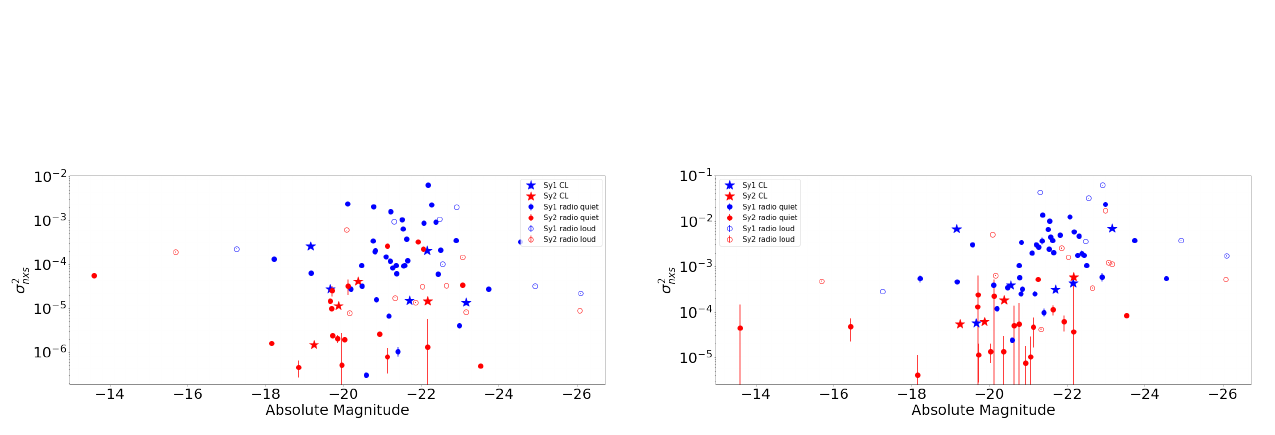}
\caption{The relationship between absolute 
magnitude and normalized excess variance for the ASAS-SN light curves (right panel) with an average uncertainty $9.0\times10^{-5}$ and TESS light curves (left panel) with an average uncertainty $6.6\times10^{-6}$. The absolute magnitude was calculated using the median V-band apparent magnitude. Radio quiet AGNs are represented with filled in circles, radio loud with open circles and CL AGNs with filled stars. 
\label{abs_mag}}
\end{figure}

To investigate and compare the variability of Sy1s and Sy2s, the relationship between the normalized excess variance and absolute magnitude was analyzed. The absolute magnitude of the sources was calculated by using their published redshifts and mean ASAS-SN magnitudes of the sources, correcting for Galactic extinction and applying the k-correction. The k-correction was calculated using the V-band AGN template found in \citet{2010ApJ...713..970A}. The Galactic extinction for each source was taken from NED with the bandpass chosen as CTIO V. The results are presented in Figure~\ref{abs_mag}. The difference in $\sigma_{\textrm{nxs}}^2$ between the TESS and ASAS-SN data is a reflection of the lengths of the corresponding light curve.

\begin{figure}[ht]
\vspace{0.3in}
\includegraphics[width=\textwidth, trim={0 0 0 3.5cm}]{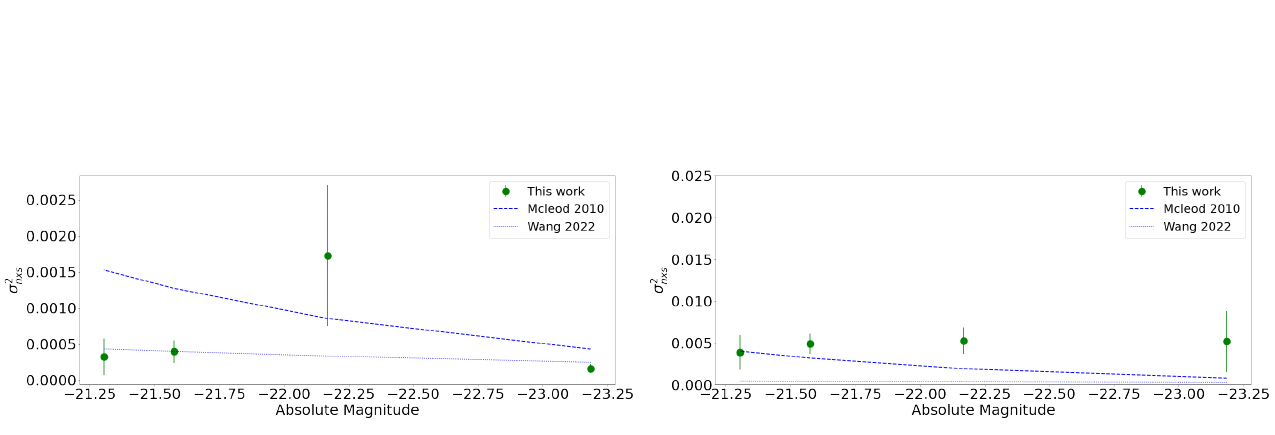}
\caption{The binned relationship of the Sy1s brighter than $-21$ V compared to the relationships found in \citet{Macleod_2010ApJ...721.1014M} in the dashed blue and \citet{wang_2022RAA....22a5014W} in the dotted blue. Left Panel: results of the binned TESS measurement, Right Panel: results of the binned ASAS-SN measurement.
\label{binned}}
\end{figure}

We binned $\sigma_{\textrm{nxs}}^2$ values for non-jetted type 1s with a magnitude brighter than $-21$~V for the TESS and ASAS-SN data in Figure~\ref{binned}, over-plotted with relations in the literature scaled to our data sets. Each bin consists of six non-jetted type 1s where the mean $\sigma_{\textrm{nxs}}^2$ of each bin is shown and the uncertainty was estimated using the standard error of the mean.
The number of sources decreases significantly below $V=-21$ suggesting the incompleteness of the sample from the parent 9-month BAT sample, and this limit corresponds to an apparent magnitude of $V=14.5$ well below the survey limits of ASAS-SN and TESS.
\citet{Macleod_2010ApJ...721.1014M} studied the trends between {SF}$_\infty$ and luminosity, redshift and black hole mass for a sample of quasars, reporting an anti-correlation between {SF}$_\infty$ and the luminosity. A re-scaling factor was applied to the relation, based on Equation~4 in \citet{Macleod_2010ApJ...721.1014M}, for TESS the case when $\Delta t \ll \tau$, for ASAS-SN the case when $\Delta t \gg \tau$, where $\tau$ is the characteristic timescale and $\Delta t$ is the length of the light curve. The difference in re-scaling the relation for the case of the TESS and ASAS-SN data sets is a consequence of the lengths of the corresponding light curves.
A conversion from variability in magnitude units to flux units was also applied. 
This relation is in excellent agreement with our measurements in this paper, especially for the accurately measured TESS data; for the larger scatter ASAS-SN measurements, the relation predicts lower $\sigma_{\textrm{nxs}}^2$ values at the luminous end.
Comparing with other relations found in the literature, for example \citet{wang_2022RAA....22a5014W} looked at how the log of the variability amplitude relates to the absolute magnitude in various bands for narrow line Sy1s and found a weak anti-correlation, their relation predicts lower $\sigma_{\textrm{nxs}}^2$ values at both the less luminous and luminous end compared to our TESS and ASAS-SN measurements. 

Although our sample size is small, we nonetheless performed a correlation test on the absolute magnitude measurements resulting in a non-statistically significant anti-correlation between the absolute magnitude and the $\sigma_{\textrm{nxs}}^2$ for the TESS and ASAS-SN data sets. 
Overall, our measurements are consistent with previously established relation of \citep{Macleod_2010ApJ...721.1014M}; however, our small sample size does not allow independent confirmation of the anti-correlation between $\sigma_{\textrm{nxs}}^2$ and luminosity due to the large intrinsic scatter of the relation.

\begin{figure}[ht]
\vspace{0.3in}
\includegraphics[width=\textwidth, trim={0 0 0 3.5cm}]{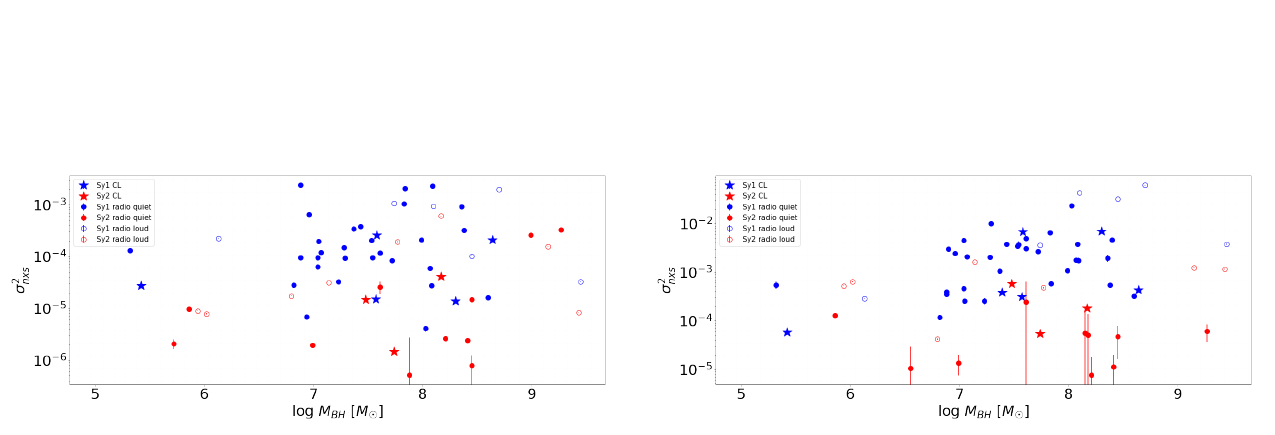}
\caption{The relationship between $M_{BH}$ and $\sigma_{\textrm{nxs}}^2$ for TESS (left panel) with an average uncertainty $3.6\times10^{-6}$ and ASAS-SN (right panel) with an average uncertainty $4.2\times10^{-5}$. Radio quiet AGNs are represented with filled in circles, radio loud with open circles and CL AGNs with filled stars.
\label{mbh}}
\end{figure}

We further examined the dependence of $\sigma_{\textrm{nxs}}^2$ with the Eddington ratio and black hole mass. The black hole masses used for calculating the Eddington luminosity for our sample were taken from the BASS project data release 2 \citep{koss_2022ApJS..261....2K}.
The V-band luminosity of our sample calculated above (Figure \ref{abs_mag}) was used in the calculation of the Eddington ratio. With the Eddington ratio being related to the accretion rate it is suggested that this is the driver behind AGN variability \citep{Macleod_2010ApJ...721.1014M}. Performing another correlation test on the black hole mass measurements and Eddington ratio measurements resulted in a non-statistically significant correlation between the black hole mass and an anti-correlation between $\sigma_{\textrm{nxs}}^2$ and the Eddington ratio for the type 1s in the TESS data set. For the ASAS-SN data set, there is a non-statistically significant correlation between the black hole mass and an anti-correlation between $\sigma_{\textrm{nxs}}^2$ and the Eddington ratio. After binning the Eddington ratio and fitting the type 1s with a power-law produced a slope of $-0.20$ $\pm$ 0.15 for TESS and 0.00 $\pm$ 0.16 for ASAS-SN. The TESS relation shows an anti-correlation compared to the ASAS-SN. Looking at the dependence of the black hole mass on $\sigma_{\textrm{nxs}}^2$ is seen in Figure~\ref{mbh}. After binning the data revealed a correlation between the black hole mass and $\sigma_{\textrm{nxs}}^2$. We fit a power-law with a slope of 3.9 $\pm$ 3.8 for the TESS data and a slope of 4.3 $\pm$  3.3 for the ASAS-SN data which is consistent with results found in \citet{Macleod_2010ApJ...721.1014M}.

\begin{figure}[ht]
\vspace{-0.3in}
\includegraphics[width=\textwidth, trim={0 0 0 2cm}]{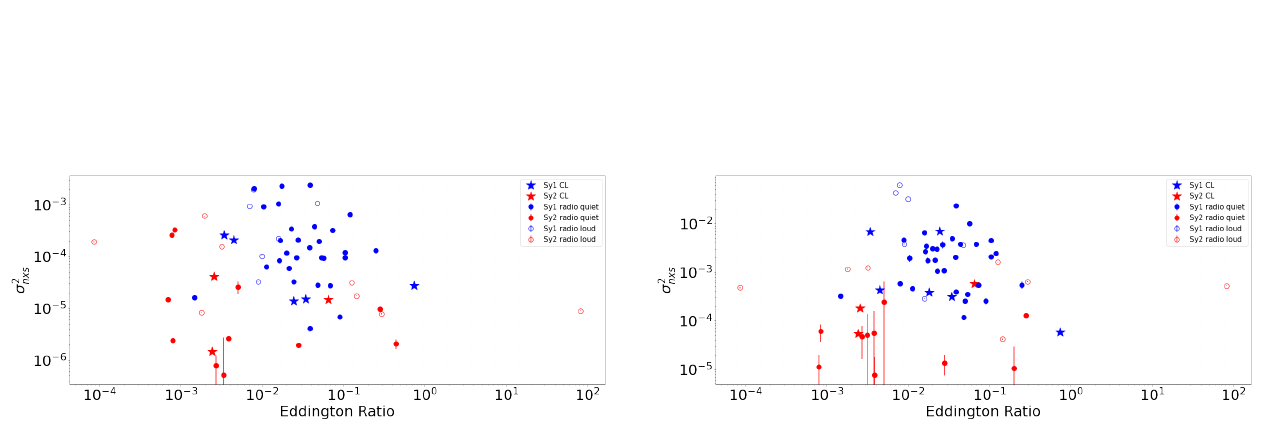}
\caption{The relationship between V-band Eddington ratio and $\sigma_{\textrm{nxs}}^2$ for the TESS (left panel) with an average uncertainty $3.6\times10^{-6}$ and ASAS-SN light curves (right panel) with an average uncertainty $4.2\times10^{-5}$. Radio quiet AGNs are represented with filled in circles, radio loud with open circles and CL AGNs with filled stars.
\label{edd_ratio}}
\end{figure}

\begin{figure}[ht]
\includegraphics[width=\textwidth, trim={0 0 0 0}]{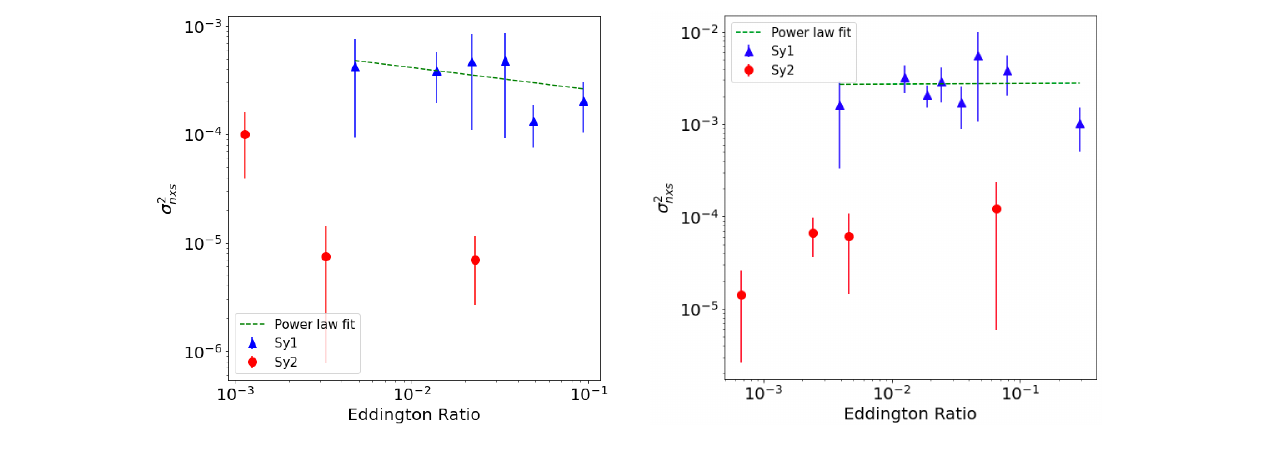}
\caption{The binned relationship between V-band Eddington ratio and $\sigma_{\textrm{nxs}}^2$. Left Panel: results of the binned TESS measurements. Right Panel: results of the binned ASAS-SN measurements. The result of our power-law fit for the type 1s is represented by the green dashed lines.
\label{binned_edd_ratio}}
\end{figure}

\begin{figure}[ht]
\vspace{0.5in}
\includegraphics[width=\textwidth, trim={0 0 0 2cm}]{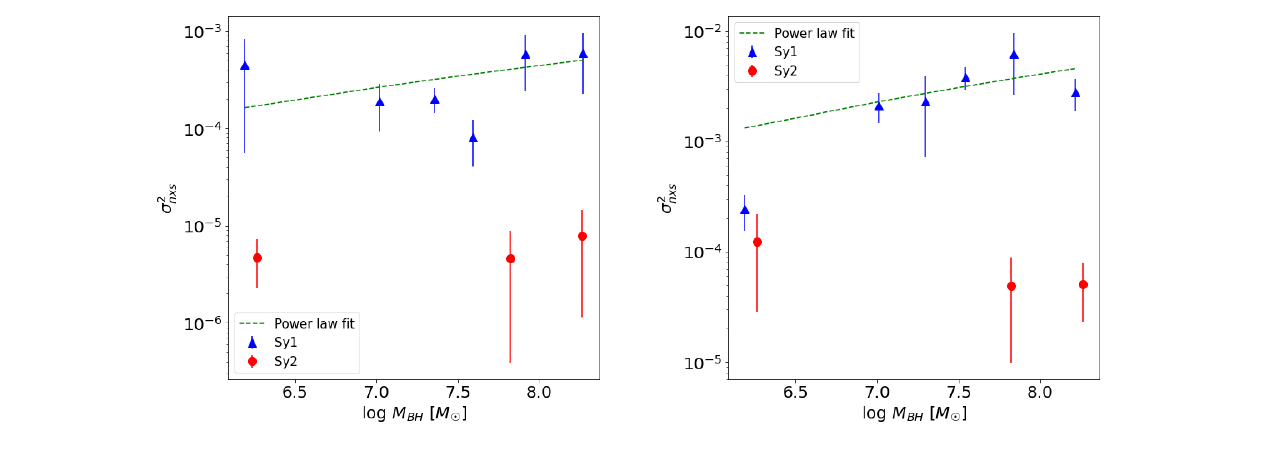}
\caption{The binned relationship between $M_{BH}$ and $\sigma_{\textrm{nxs}}^2$. Left Panel: results of the binned TESS measurements. Right Panel: results of the binned ASAS-SN measurements. The result of our power-law fit for the type 1s is represented by the green dashed lines.
\label{binned_mbh}}
\end{figure}

We calculated the average values for $\sigma_{\textrm{nxs}}^2$ for the Sy1 and Sy2 populations.  Using the ASAS-SN data set, the averages are $(3.04$ $\pm$ $0.14)\times10^{-3}$ and $(4.5$ $\pm$ $4.7)\times10^{-5}$ respectively, and using the TESS data set the average $\sigma_{\textrm{nxs}}^2$ are $(5.10$ $\pm$ $0.63)\times10^{-4}$ and $(2.6$ $\pm$ $6.3)\times10^{-5}$ respectively. The means of the Sy1 sample of either the ASAS-SN or TESS data are significantly higher than the mean values of Sy2s.
We treated any source that produced a negative $\sigma_{\textrm{nxs}}^2$ as having zero variability.  This can lead to biases when analyzing the binned or averaged data.  Only a few Sy1s have negative $\sigma_{\textrm{nxs}}^2$ values, and they are all close to zero.  We have tested that treating them as negative values has a minimal effect in our binned analysis of Sy1 variability relations in this section. For the Sy2 population, 35\% of the sample exhibits a negative variability. Inclusion of these negative values changes the result of the average $\sigma_{\textrm{nxs}}^2$ for both the TESS and ASAS-SN data sets. For TESS the average is decreased to $(1.8$ $\pm$ $6.3)\times10^{-5}$, for ASAS-SN the average decreases to a negative value of $(-3.4$ $\pm$ $1.8)\times10^{-4}$.

\subsection{Kolmogorov-Smirnov test} \label{sec:KS}
\begin{figure}[ht]
\includegraphics[width=1.0\textwidth, trim={0 0 0 1cm}]{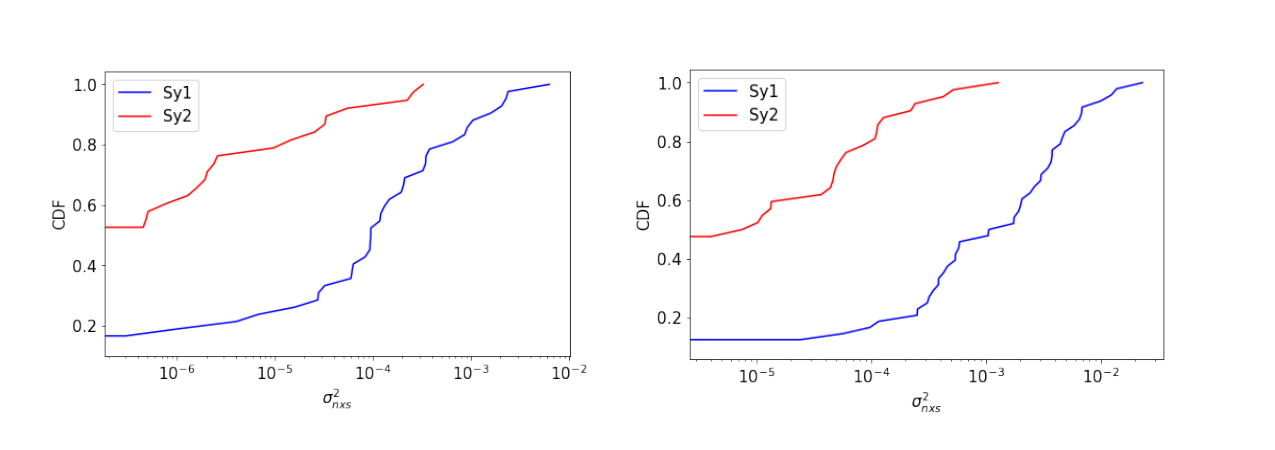}
\caption{The cumulative distribution function of the normalized excess variance of Sy1 and Sy2 galaxies. Types 1, 1.2 and 1.5 are placed in the same sample and are shown in blue. The types 2, 1.8 and 1.9 are placed in the same sample and are shown in red. Radio loud sources are not included. Left panel: presents the results from TESS measurements. Right panel: the results from ASAS-SN measurements. Both results signify a clear distinction in the variability between the two samples. 
\label{KS_test}}
\end{figure}

To evaluate the difference in the distribution of the variability between the non-jetted type 1 and type 2 AGNs a two-sample Kolmogorov-Smirnov (KS) test was performed. The KS test was first performed on the TESS and ASAS-SN datasets separately, yielding p-values of $2.4\times10^{-7}$ and $1.4\times10^{-12}$, respectively, suggesting that Sy1s and Sy2s exhibit significantly different variability amplitudes. Figure~\ref{KS_test} shows the cumulative distributions of $\sigma_{\textrm{nxs}}^2$ values for the type 1 and 2 samples using ASAS-SN and TESS data, where the distributions can be visually distinguished. 
We further test the variability difference between type 1s and type 2s by separating the total sample into two bins 
by either the absolute magnitude, Eddington ratio, or black hole mass.  The KS tests were performed on the variability of Seyfert 1s and 2s in each bin for the TESS and ASAS-SN data sets separately. Results of these tests range in p-values from 0.04 -- $1.3\times10^{-5}$ for TESS and $6.5\times10^{-4}$ -- $3.0\times10^{-10}$ for ASAS-SN, with median values of $6\times10^{-4}$ and $3.3\times10^{-5}$ respectively. These results show that the variability difference between type 1s and 2s is always significant, regardless of the AGN parameters in consideration, such as the luminosity, Eddington ratio, or black hole mass.

This is consistent with the results seen in \citet{2023MNRAS.518.1531L} comparing the variability of weak Sy1s and 2s, classifying a weak type 1 as having broad lines detected at a lower significance ($<$5$\sigma$) than required for the SDSS pipeline, who reported that there is a distinction in the variability between even weak Sy1s and 2s based on the KS test and claiming that the variance of the type 2s are smaller by about an order of magnitude than the type 1s. 

Other KS tests were performed; this includes on the jetted and non-jetted type 1s in the TESS and ASAS-SN datasets and the jetted and non-jetted type 2s in the TESS and ASAS-SN datasets; yielding p-values of 0.64, 0.17, $4.75\times10^{-5}$ and $1.19\times10^{-7}$ respectively. These KS test results are limited by the number of jetted AGNs in our sample, such that the distribution of variability amplitude between jetted and non-jetted AGNs is not significant. However, type 2 jetted and non-jetted AGNs are distinguished.

\subsection{Seyfert 2s with non-negligible short time-scale variability}
Majority of the Sy2s in this sample have a negligible or near negligible variability which is consistent with predictions from the unification model (\citealt{Antonucci_1993ARA&A..31..473A,Urry_1995PASP..107..803U}). In another work by \citet{Yip_2009AJ....137.5120Y} found no evidence of spectral variability for type 2s in the Sloan Digital Sky Survey. 
Among the non-jetted Sy2s, a fraction of them have significantly detected $\sigma_{\textrm{nxs}}^2$ above zero value, and their variability is manifested in the very long timescale spanning multiple years.  
Some of the Sy2s in our sample that do show relatively large $\sigma_{\textrm{nxs}}^2$ are classified as radio loud sources making their variability most likely the result of the jet and not the accretion disk itself. 

\begin{figure}[ht]
\includegraphics[width=1.0\textwidth, trim={0 0 0 1.6cm}]{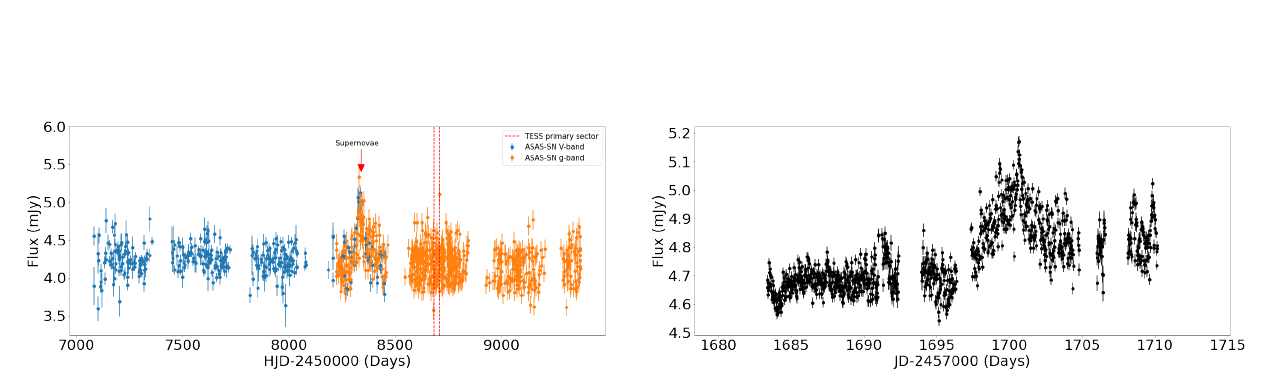}
\caption{Left: ASAS-SN light curve in both V-band and g-band (the V and g light curves are normalized) of the Sy2 MCG+04-48-002. A non-stochastic variability in the form of a flare is seen in the middle of the light curve, which coincides with a previously reported supernova explosion in the galaxy. The red dotted lines indicate the time frame the TESS light curve takes place. Right: TESS light curve that takes place after the flare exhibiting unusual variability.
\label{mcg0448022}}
\end{figure}

The few that are classified as radio-quiet and also show small variations are of interest to further examination.  One type 2 with non-negligible variability is MGC+04-48-002, which is part of the dual AGN system SWIFT J2028.5+2543 that also includes the Sy2 NGC 6921 \citep{2016ApJ...824L...4K}, separated by 25.3 kpc (91''). However, with the resolution of both TESS and ASAS-SN, we are able to resolve the two AGNs therefore the light curve presented is that for MGC+04-48-002. We can see that in the TESS and ASAS-SN light curves in Figure~\ref{mcg0448022} there is variability in both light curves. In the ASAS-SN light curve, we can see a non-stochastic variability in the form of a flare in the middle of the light curve. An archival search on the radio properties of this source was conducted showing no evidence of the AGN having a radio jet, ruling that out as a potential cause of the flare. \citet{stritzinger_2018TNSCR2131....1S} reported a supernova in the host galaxy at the flare time which would explain the variability seen. With the flare being the result of a supernova it was filtered out of the light curve and the normalized excess variance was recalculated. However, examining the TESS light curve that is after the supernova, we still see short timescale variability different from that of a normal Sy2, making this source interesting for further follow-up.

\subsection{Changing Look AGNs} \label{sec:CL}
The \textit{Swift}-BAT sample contains several well studied and previously identified CL AGNs. These include NGC 4151, 3C390, Fairall 9, Mrk 590, NGC 2992, Mrk 1018, NGC 3516, NGC 7582, NGC 526A, (\citealt{Malkov_1997A&A...324..904M,  Penston_1984MNRAS.211P..33P,Wamsteker_1985ApJ...295L..33W, Denney_2014ApJ...796..134D,gilli_2000A&A...355..485G,cohen_1986ApJ...311..135C,Shapovalova_2019MNRAS.485.4790S}) and a new {\textquotedblleft turn off\textquotedblright} event in the CL AGN NGC 1365 \citep{temple_2023MNRAS.518.2938T}. These CL AGNs account for eight out of the 55 Sy1s and four out of 53 Sy2s in our sample.
A few of these CL events took place during the time frame of the ASAS-SN light curves. For those CL AGNs, we can study the light curve characteristics alongside the transitions reported in the literature. 

\subsubsection{NGC 1365}
\begin{figure}[ht]
\includegraphics[width=1.0\textwidth, trim={0 0 0 1.5cm}]{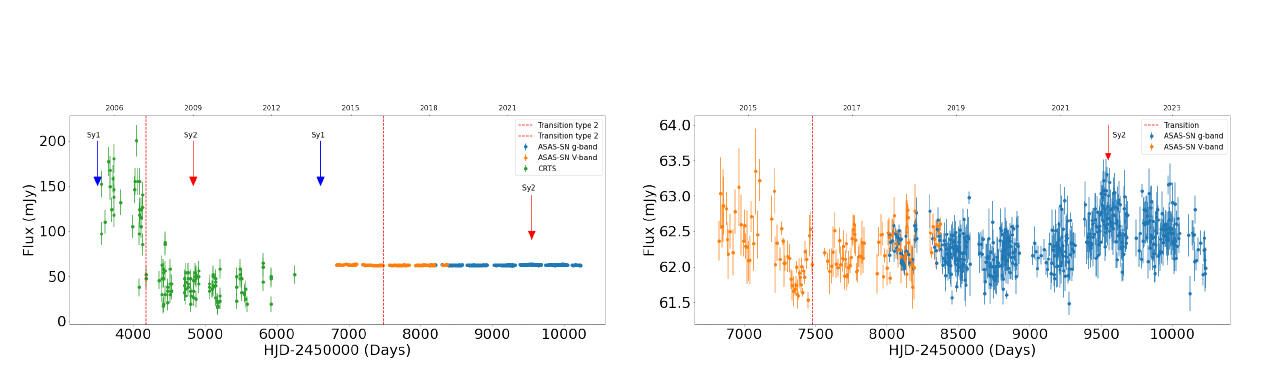}
\caption{CL AGN NGC 1365, changed from a type 1 to type 2. Right: displaying the change in the ASAS-SN light curve. Left: The CTRS light curve before this change shows a previous type change. Red vertical dotted lines represent transition times based on the light curves, and the downward arrows mark the spectroscopic observations with classifications.
\label{ngc1365}}
\end{figure}

NGC 1365 has a history of transitioning types multiple times. \citet{Schulz_1999A&A...346..764S} reports in 1999 a broad H$\beta$ component that has a ratio of broad/narrow close to that reported in an earlier paper for H$\alpha$, suggesting that in 1999 NGC 1365 had a broad H$\beta$ and H$\alpha$ classifying it a Sy1. However, a 2009 January spectrum taken by \citet{Trippe_2009PhDT.......121T} shows little to no evidence of a broad H$\alpha$ and reports a narrow H$\beta$ classifying it as a type 1.9, signifying that the AGN transitioned from a Sy1 to a Sy2 from 1999--2009. We can see this transition in the CRTS light curve \citep{Drake_2009} presented in the left panel of Figure~\ref{ngc1365}, where the first part of the light curve is more variable and then dips and flattens out. Spectra taken in 2013 December show broad Balmer lines meaning that the source transitioned back to a type 1 \citep{temple_2023MNRAS.518.2938T}. The source was re-observed in 2021 December with Magellan MagE and \citet{temple_2023MNRAS.518.2938T} reported finding only narrow H$\beta$ and H$\alpha$, stating that this is a new CL event for NGC 1365 where it has transitioned back to a Sy2. This transition can be visually observed in the ASAS-SN light curve in Figure~\ref{ngc1365} (right). The initial part of the light curve is visually more variable suggesting a type 1 classification. However, the latter part of the light curve can be seen as showing less variability making it that of a type 2.

We can estimate the type transition time by moving the transition boundary in the window between the two spectroscopic observations that signifying a type transition, such that 
the variability difference  before and after the boundary is maximized.
Although this estimation from the light curve is associated with uncertainties, it can better constrain this transition time compared to more sparsely separated spectroscopic observations.

\subsubsection{Mrk 590} 
\begin{figure}[ht]
\vspace{0.6in}
\includegraphics[width=1.0\textwidth, trim={0 0 0 3.5cm}]{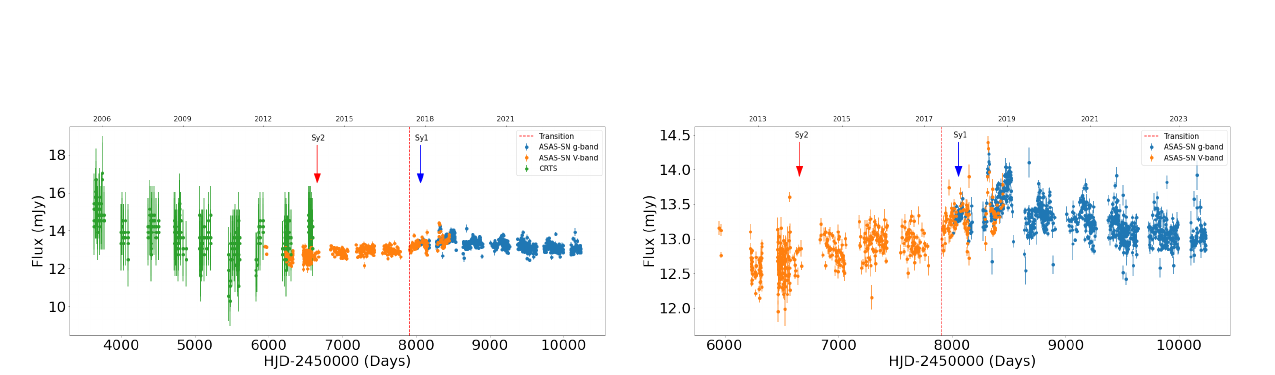}
\caption{CL AGN Mrk 590, changed from a type 2 to type 1. Left: displaying the change in the ASAS-SN light curve. Right: The CTRS light curve before this change shows when the AGN was classified as a type 2. Red vertical dotted lines represent transition times based on the light curves, and the downward arrows mark the spectroscopic observations with classifications.
\label{mrk590}}
\end{figure}

The CL AGN Mrk 590 has been studied thoroughly and it was the object of a more than 40-year multi-wavelength study in \citet{Denney_2014ApJ...796..134D}. In this study, it shows that Mrk 590 transitioned from a typical Sy1 to a Sy2. This is seen in the brightening of the H$\beta$ reaching its peak in 1989 and then dimming to being completely gone by 2006. There is also no broad H$\beta$ line in 2013 February, 2013 December, and 2014 January. In MUSE observations taken in October and November of 2017 \citep{Raimundo_2019MNRAS.486..123R} shows the reappearance of both the broad H$\beta$ and a broad H$\alpha$ line, meaning that the AGN transitioned back from a type 2 to a type 1 between 2014 January and 2017 October. In Figure~\ref{mrk590} we see that the first half of the ASAS-SN light curve for Mrk 590 resembles that of a type 2, while the second half is more variable similar to that of a type 1. This change in the variability of the light curve takes place in roughly 2017 June. Lining up with the timeline based on spectral observations, meaning this is most likely when the AGN transitioned from a type 2 to type 1. We also included the CRTS light curve \citep{Drake_2009} in the left panel of Figure~\ref{mrk590} to confirm that it resembles that of a Sy2 as previously reported.

\subsubsection{NGC 3516}
\begin{figure}[ht]
\vspace{0.9in}
\includegraphics[width=1.07\textwidth, trim={1cm 0 0 3.8cm}]{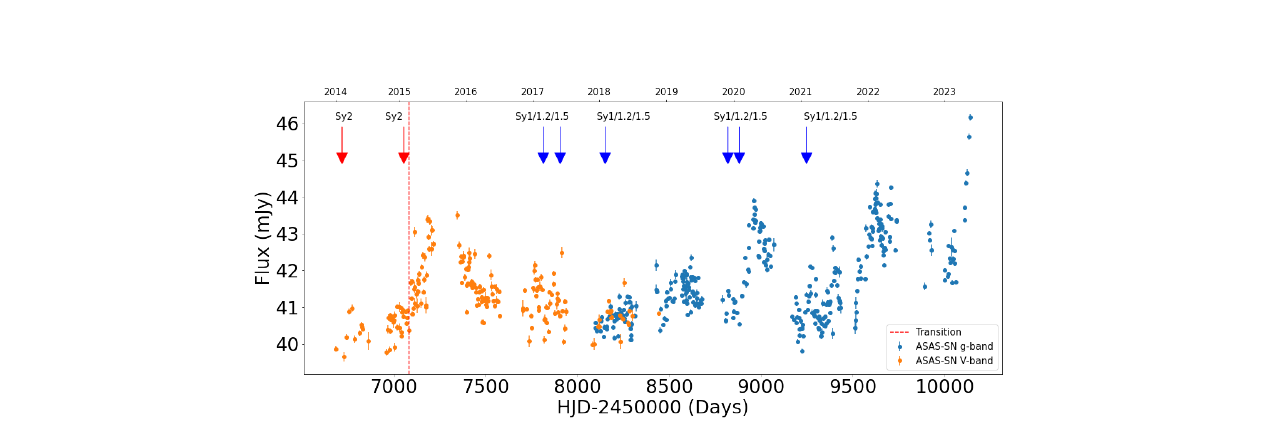}
\caption{CL AGN NGC 3516, changed from a type 2 to type 1 in the ASAS-SN light curve. Red vertical dotted lines represent transition times based on the light curves, and the downward arrows mark the spectroscopic observations with classifications.
\label{ngc3516}}
\end{figure}

NGC 3516 has undergone a long-term optical spectral monitoring campaign presented in \citet{Shapovalova_2019MNRAS.485.4790S}, where it was shown to transition from a Sy1 to a Sy2. In spectral observations that were taken in this study in 2007, NGC 3516 presented prominent broad Balmer lines specifically H$\alpha$ and H$\beta$, classifying it as a typical Sy1. However, in the 2014 spectra, no broad lines were present and only a narrow H$\alpha$ is seen making the new classification a Sy2. This means that at the start of our ASAS-SN light curve, the AGN was a type 2, visually in the first part of the light curve there is not a lot of variability. While examining the later part of the light curve we see increased variability with multiple peaks starting in 2019. In \citet{Popovi_2023A&A...675A.178P} a companion paper to \citet{Shapovalova_2019MNRAS.485.4790S} states that a very weak broad H$\alpha$ and H$\beta$ were detected in their 2017 to 2020 spectra and that the intensity of the H$\beta$ line is increasing as seen in their 2021 spectra. The emergence of the broad Balmer lines would explain the change and gradual increase in the variability seen in the later part of the light curve, resembling that of a type 1. Based on the change in the light curve this transition back to a type 1 can be seen and better pinpointed as the dashed red line in Figure~\ref{ngc3516}.

\begin{figure}[ht]
\vspace{1.1in}
\includegraphics[width=1\textwidth, trim={1cm 0 0 3.5cm}]{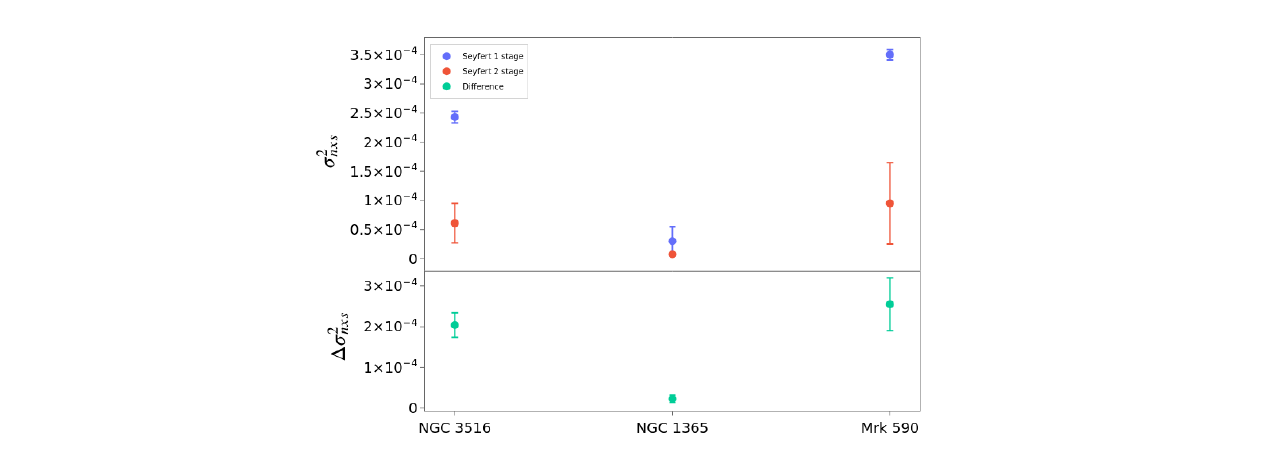}
\caption{Top panel: The range of variability resulting from moving the transition line between the spectroscopic observation window.  The data points are calculated at the estimated transition boundary, which gives the maximum $\sigma_{\textrm{nxs}}^2$ difference and the 
 uncertainties represent the range of $\sigma_{\textrm{nxs}}^2$ values calculated in the transition window. 
 The variability when the CL AGN is in a type 1 stage is always larger than when it is in a type 2 stage.
Bottom Panel: The difference of the $\sigma_{\textrm{nxs}}^2$ at the estimate transition time.
\label{sum}}
\end{figure}

We compiled the normalized excess variance from before and after the AGN changed types and displayed the results in Figure~\ref{sum}. 
The uncertainties indicate the range of the differences between the normalized excess variance in the window between the spectroscopic observations, which are much larger than the uncertainties of the normalized excess variance difference at the estimated transition boundary.
Here again, it is shown that when the AGN was/is a type 2 it displayed lower variability than when it was/is classified as a type 1. 
This result does not depend on the exact estimated transition time, but applies to any transition boundary in the transition window between the two spectroscopic observations. 
This change of variability can potentially be a viable option for determining CL candidates, which can also be seen in the works by \citet{Yang_2018ApJ...862..109Y}.

\section{Discussion and Conclusions} \label{conclusion}
We analyzed the ASAS-SN and TESS light curves for a sample of Sy1s and Sy2s from the \textit{Swift}-BAT 9-month survey. In our analysis, we compared the long-term (ASAS-SN) and short-term (TESS) variability of the two types of AGNs in an effort to determine a clear distinction in the variability between the two types. We measured the variability by the normalized excess variance and constrained how their variability compares with their luminosity, Eddington ratio, and black hole mass. For the absolute magnitude, we found that for the brighter type 1s ($<$ $-$21) there is a weak anti-correlation between the variability and luminosity for the TESS measurements and the ASAS-SN measurements. This matches previous works done by \citet{Macleod_2010ApJ...721.1014M} and  \citet{wang_2022RAA....22a5014W}. We also examined how the Eddington ratio and black hole mass depend on the normalized excess variance showing an anti-correlation with the Eddington ratio and a correlation with the black hole mass which is again consistent with results from \citet{Macleod_2010ApJ...721.1014M}. AGN variability has been extensively studied in both the optical and X-ray bands, which probe the accretion disk and corona components of the central engine.  Although the disk and corona are related by the inverse Compton scattering and reprocessing processes, where the UV seed photons are up scattered by energetic electrons in the corona to produce the X-ray emission and X-rays are irradiating the disk during reprocessing and propagating X-ray variability to the optical band, the connection between optical and X-ray variability is not simple. Both the X-ray and optical PSD breaks were found to correlate with black hole mass \citep[e.g.,][]{Kelly_2009, Macleod_2010ApJ...721.1014M, Gonz_2012A&A...544A..80G, Gonz_2018ApJ...858....2G, burke_2021Sci...373..789B}, but the breaks are at significantly different time scales ($10^{-3}$ to $10^{-2}$ day$^{-1}$ for optical and $10^{0}$ to $10^{2}$ day$^{-1}$ for X-ray) and they scale with the black hole mass with different slopes--about $-2.5$ for optical \citep{burke_2021Sci...373..789B}, but about $-1.0$ for X-ray \citep{Gonz_2012A&A...544A..80G} for the black hole mass to break frequency relation.
\citet{Yuk_2023arXiv230617334Y} found a new set of high-frequency optical breaks at $10^{-2}$ to $10^{-1}$ day$^{-1}$ frequencies, which better correlates with the X-ray breaks, supporting an overall underlying physical link between optical and X-ray variability, further corroborated by continuum reverberation mapping studies (\citealt{edelson_2015ApJ...806..129E,mchardy_2018MNRAS.480.2881M}).
Comparing the variability amplitude scaling relations, as described by the relation between the normalized excess variance and other AGN parameters, between optical and X-ray bands, 
the optical correlations have weaker dependencies on other AGN parameters, i.e the Eddington ratio or black hole mass. The optical correlations being $-0.23 \pm 0.03$ and $0.11 \pm 0.02$ respectively \citep{Macleod_2010ApJ...721.1014M}, in contrast to the X-ray correlations $1.11 \pm 0.13$ and $-0.91 \pm 0.09$ \citep{Ponti_2012A&A...542A..83P} respectively.
The optical correlation with luminosity presented in this paper is negative, which is consistent with the trend in the X-ray band.
However, the optical correlation with black hole mass is positive and with Eddington ratio negative in contrast to the measurements in the X-ray band, where there exists an anti-correlation between the variability and black hole mass and potential positive correlation with Eddington ratio (\citealt{Ponti_2012A&A...542A..83P,akylas_2022A&A...666A.127A}). 
It is interesting to note that $L \propto \lambda M_{BH}$, where $\lambda$ is the Eddington ratio.
Both the optical and X-ray correlations with Eddington ratios are weak and can be argued as not significant.  
The energy bands presented in the X-ray analysis of \citet{akylas_2022A&A...666A.127A} are the $3-10$ keV and $10-20$ keV bands with an average duration for the \textit{NuSTAR} observations of approximately 80 ks. In contrast, the optical timescales of the ASAS-SN and TESS light curves span roughly 10 years and 27 days respectively, at longer time scales compared to those X-ray studies. It would be beneficial to constrain these relations on longer time scales in the X-ray band and even shorter time scales in the optical band for a complete picture of AGN variability amplitude dependency on AGN parameters. 
The combined optical and X-ray results present a complex picture of AGN variability, which will provide constraints and challenges to AGN theoretical models.

A KS test on the non-jetted type 1 and type 2s revealed that type 1s are more variable in both the short and long timescales. This was compared to other studies that showed similar results \citep{2023MNRAS.518.1531L}. 
The Sy1 population also shows a much higher mean normalized excess variance value than Sy2s.  Based on the AGN unification model, the difference between the Sy1s and 2s is due to the presence of a torus in the line of sight to observers.  Thus, this variability difference between the Sy1 and Sy2 populations could be interpreted under the unification scheme as scattering from the large-scale dusty torus region, suppressing the Sy2 variability. X-ray data also support this, \citet{papadakis_2024A&A...685A..50P} reported using $14-195$ keV band light curves that the PSD’s of Sy1s and Sy2s are identical, again in agreement with the unification model, due to the hard X-rays ability to pierce through the dusty torus and retain the variability of Sy2s.

Further testing the variability difference between type 1 and 2s in luminosity, Eddington ratio, and black hole mass sub-samples, we found that the variability of type 1s are always significantly larger than that of type 2s, regardless of the detailed AGN parameter range in consideration. 
This supports the unification models by the difference between the line-of-sights of type 1s and 2s, rather than by time, where type 2s are considered at the early stage of AGN evolution enshrouded by the obscuring dust \citep[e.g.,][]{Granato_2004ApJ...600..580G}.
 
Within the parent sample, there is a small subsample of changing-look AGNs that we analyzed. In particular, NGC 1365, Mrk 590 and NGC 3516 were reported to have transitioned during the time frame of the ASAS-SN light curves. Through calculations of the variability of the light curves before and after a moving line in the window between spectral observations, we estimated when the transitions occurred. Suggesting that a combination of photometric and spectroscopic monitoring campaigns can more precisely determine the transition time.
It is interesting to observe that the senses of variability difference between non-changing Seyfert 1s and 2s and CL AGNs in 1 or 2 stages are the same.
This consistency trend suggests that variability can be a key factor in understanding the changing-look AGNs or the dichotomy between Seyfert 1 or 2 populations.

\begin{acknowledgments}
We thank the constructive comments from the anonymous referee and C.~S.\ Kochanek. 
 N.~K., X.~D., and H.~Y.\ would like to acknowledge NASA funds
80NSSC22K0488, 80NSSC23K0379 and NSF fund AAG2307802. We thank Las Cumbres Observatory and its staff for their continued support of ASAS-SN. ASAS-SN is funded by Gordon and Betty Moore Foundation grants GBMF5490 and GBMF10501 and  the Alfred P. Sloan Foundation grant G-2021-14192. K.Z.S is supported by NSF grants 2307385 and 2407206. 
\end{acknowledgments}

\vspace{5mm}
\facilities{ASAS-SN, TESS, \textit{Swift}}

\bibliography{bibliography}{}
\bibliographystyle{aasjournal}

\end{document}